\documentclass{article}%
 
\usepackage{graphicx}
\usepackage{cite}
\usepackage{amsmath,amssymb,amsfonts}
\usepackage{algorithm}
\usepackage{algpseudocode}
\usepackage{graphicx}
\usepackage{textcomp}
\usepackage{xcolor}
\usepackage{hyperref}
\usepackage[capitalise]{cleveref}

\usepackage{tabularx}
\usepackage{adjustbox}
\usepackage{amsthm}

\newtheorem{theorem}{Theorem}
\newtheorem{definition}[theorem]{Definition}

\newtheorem{lemma}[theorem]{Lemma}
\newtheorem{corollary}[theorem]{Corollary}

\DeclareMathOperator{\EX}{\mathbb{E}}
\DeclareMathOperator{\RK}{\text{rank}}

\newcommand{\bigo}{\mathcal{O}}
\newcommand{\softo}{\Tilde{\mathcal{O}}}

\usepackage{geometry} 

\begin{document}

\title{Fast Order Statistics with Group Inequality Testing}
\author{Adiesha Liyanage\footnote{Gianforte School of Computing, Montana State University, Bozeman, MT 59717, USA.
Email: {\tt a.liyanaralalage@montana.edu}}
\and
Brendan Mumey\footnote{Gianforte School of Computing, Montana State University, Bozeman, MT 59717, USA.
Email: {\tt brendan.mumey@montana.edu}}
\and
Braeden Sopp\footnote{Gianforte School of Computing, Montana State University, Bozeman, MT 59717, USA.
Email: {\tt braeden.sopp@student.montana.edu}}}

\date{}
\maketitle

\begin{abstract}
Suppose that a group test operation is available for checking order relations in a set, can this speed up problems like finding the minimum/maximum element, determining the rank of element, and computing order statistics? We consider
a one-sided group inequality test to be available, where queries are of the form $u \le_Q V$ or $V \le_Q u$, and the answer is `yes' if and only if there is some $v \in V$ such that $u \le v$ or $v \le u$, respectively.
We restrict attention to total orders and focus on query-complexity; for min or max finding, we give a Las Vegas algorithm that makes $\bigo(\log^2 n)$ expected queries.
We observe that rank determination can be solved with existing ``defect-counting'' algorithms, but also give 
a simple Monte Carlo approximation algorithm with expected query complexity 
$\softo(\frac{1}{\delta^2} \log \frac{1}{\epsilon})$, where $1-\epsilon$ is the probability that the algorithm succeeds and we allow a relative error of $1 \pm \delta$ for $\delta > 0$ in the estimated rank. 
We then give a Monte Carlo algorithm for approximate selection that has expected
query complexity $\softo(\frac{1}{\delta^4}\log \frac{1}{\epsilon \delta^2} )$; it has probability at
least $\frac{1}{2}$ to output an element $x$, and if so, $x$ has the desired approximate rank with
probability $1-\epsilon$.

\end{abstract}

\section{Introduction}

In this work, we consider how a group inequality test can speed up order-finding problems, such
as finding the minimum/maximum element of a totally-ordered set, determining the rank of an element, and selecting the element with a given rank.  
Given an order relation $R$ on a set $U$ of size $n$, we assume that an oracle is available that can answer 
one-to-many queries of the form $u \le_Q V$ or $V \le_Q u$,
where the answer is yes if and only if there is some $v \in V$ that satisfies the inequality $u \le v$ or $v \leq u$ respectively.

As a motivating example, imagine that we are surveying someone to determine their preferred choice  among a set of options, but we can only ask yes or no questions.  In this case the group test queries proposed are natural, i.e. do you prefer A versus B or C?  
Another example where such a test is available occurs in computational biology, where one seeks to identify consistent relationships among all parsimonious solutions for reconstructing tumor cell evolution from single cell sequence data~\cite{Liyanage2025.04.12.648524}. 
In this case, the goal is to determine those order relationships among pairs of cells that hold true in all optimal phylogenetic trees on the cells. In \cite{Liyanage2025.04.12.648524}, the authors first present a group test $u \le_Q V$ based on integer-linear programming that can check whether cell $u$ evolutionarily precedes some cell $v \in V$, then develop an 
algorithm based on the test to determine the complete relationship among all cells.
In such contexts, if the goal is to find certain elements (e.g., the minimum or maximum), it is natural to ask whether group testing strategies can be effectively applied. 


\subsection{Related Work}

Selection is a fundamental and well-studied problem in combinatorics and computer science; given an order relation \(R\) on a set \(U\) of size \(n\), the task is to find the \(k\)-th smallest element.
In the traditional comparison model, deterministic algorithms exist that can select the \(k\)-th element using at most \(\bigo(n)\) comparisons~\cite{blum1973time},
while a randomized selection algorithm exists that further lowers the number of comparisons to \(n + min(k, n-k) + o(n)\) in the average case.
In large-scale data applications, the space required for exact selection algorithms can be impractical, so approximate selection algorithms have been developed with a focus on space efficiency.
Approximate selection algorithms are closely related to approximate quantile computation, as studied in~\cite{manku1998approximate}. Both aim to identify items whose ranks are near a target value. The key difference lies in their input specification: selection algorithms target an element with a specified integer rank, whereas quantile computation seeks an item whose rank corresponds to a given fraction (quantile) of the total input size. The quantile of an item~$x$ is defined as the fraction of items in the stream satisfying $x_i \leq x$.
Approximate quantile computation algorithms such as the deterministic and randomized one pass algorithms presented in~\cite{manku1998approximate} achieve space complexity \(\bigo(\frac{1}{\epsilon}\log^2 \epsilon n)\). 
The current state-of-the-art algorithm for approximate quantiles computation in a streaming setting, such as the KLL sketch~\cite{karnin2016optimal}, achieves space complexity of \(\bigo(\frac{1}{\delta} \log\log \frac{1}{\epsilon})\) for an additive error of \(\delta n\) with success probability \(1-\epsilon\). 

Moreover, the group testing literature contains extensive work on estimating the number of defective items in a population
\cite{swallow1985group, chen1990using, thompson1962estimation, walter1980estimation, gastwirth1989estimation}.
In this setting, let \(X\) denote a set of elements and let \(D \subseteq X\) be an unknown subset of defective elements.
We are given access to an oracle that, on input a subset \(T \subseteq X\), returns `true' if and only if \(T\) contains
at least one defective element. The objective is to estimate the cardinality \(d = |D|\).
Two main classes of algorithms have been studied in the literature: \emph{non-adaptive} algorithms, in which all queries are fixed
in advance, and \emph{adaptive} algorithms, in which future queries can depend on the outcomes of previous queries.
In \cite{falahatgar2016estimating}, the authors proposed an adaptive randomized group testing algorithm that estimates the number
of defective elements using a near-optimal number of queries. Specifically, their algorithm uses at most
\(2 \log \log d + \bigo\left(\frac{1}{\delta^2}\log\frac{1}{\epsilon}\right)\)
queries and outputs an estimate of \(d\) within a multiplicative factor of \(1 \pm \delta\) with error probability at most
\(\epsilon\). Furthermore, they established an information-theoretic lower bound showing that any adaptive algorithm must make at
least \((1-\epsilon)\log \log d - 1\) queries to estimate \(d\) for constant \(\delta\).


In this work, we propose adaptive algorithms for selection and rank determination, assuming an oracle is available that can answer group inequality tests. We focus on minimizing the query complexity.

\subsection{Contributions}
Our results can be summarized as follows:

\begin{itemize}
    \item When considering min-finding,
    we introduce a Las Vegas algorithm to find the minimum element of a totally-ordered set \(U\) using an $\bigo(\log^2 n)$ expected group test queries
    where \(n\) is the size of \(U\). The min-finding algorithm can also be adapted for max-finding (by just reversing the group tests).

    \item  
    While we observe that rank-estimation can be solved with defect-counting group testing algorithm, we also give a simple
    Monte Carlo algorithm to estimate \(r\) for the rank of an element \(x \in U\) is computable by a  utilizing \(\bigo(\frac{\log n}{\delta^2} (\log \log n + \log \frac{1}{\epsilon}))\) expected group test queries for any given
    \(\varepsilon\) probability of failure and approximation parameter \(\delta \in (0,1)\).
    The algorithm guarantees that \(|rank(x)-r| < \delta \min(r, n-r)\) with  probability $1 - \epsilon$.
    \item For approximate selection,
     we provide a Monte Carlo algorithm that uses 
     \(\bigo(\frac{1}{\delta^2}\log^2 \frac{n}{k} + \frac{1}{\delta^4}\log \frac{1}{\epsilon \delta^2})\) expected group test queries for a given \(\delta \in (0,1)\) and target rank \(k\) which returns an element \(x \in U\) with probability 1/2. When the algorithm returns an element, then with probability \(1-\epsilon\),
     we have \(|\RK(x) - k| < \delta \min(k, n-k)\).
\end{itemize}
To the best of our knowledge, this work is the first to consider the problems of rank selection, min/max finding, and rank determination in a group testing framework.

\section{Preliminaries}

We assume that a total order $(S,\le)$ is given, where $S$ is a set of elements stored in an array or similar data structure.  The algorithms make use of random sampling and partitioning in $S$, but we will not consider the time complexity of those operations, as we focus on the query-complexity, i.e., the number of group inequality tests performed.  We assume that two group test operations are available, $u \le_Q V$ and $V \le_Q u$, where the answer is yes if and only if there is some $v \in V$ that satisfies the inequality.

\begin{definition}
For any subset $V \subseteq S$ and element $x \in S$, a left (right) group test operation is of the form  $x \le_Q V$ ($V \le_Q x$), where the test returns true if and only if there is some $v \in V$ that satisfies $x \le v$ ($v \le x$).
\end{definition}

\begin{definition}
The rank of an element $x \in S$ is its order position, i.e.,
$$
\RK(x) = |\{y \in S : y \le x\}|.
$$
\end{definition}

\begin{definition}
The $k$-th order statistic (for $k \in [n]$) is the element $x \in S$ such that $\RK(x) = k$.  Given $k$, the problem of finding the $k$-th order statistic is called the \emph{selection problem}.
\end{definition}
We first consider the min-finding problem, i.e., computing the $1$-st order statistic, and
develop an exact group testing algorithm for this problem.  
Let $\le^{\text{rev}}$ be the reversed relation, i.e., 
$x \le^{\text{rev}} y \Leftrightarrow y \le x$ (clearly $\le^{\text{rev}}$ is also a total order).  
Thus, max-finding for $\le$ is the same problem as
min-finding for $\le^{\text{rev}}$.  Furthermore, 
$x \le^{\text{rev}}_Q V$ if and only if $V \le_Q x$, and similarly, 
$V \le^{\text{rev}}_Q x$ if and only if $x \le_Q V$.  This means that we can convert a min-finding
algorithm to max-finding by simply reversing the direction of each group test call.

For the rank determination and
selection problems, we consider approximate solutions, with the following problem
definitions:
\begin{definition}
A $\delta$-approximation of $\RK(x)$ (for $\delta \in (0,1), x \in S$) is a value $r \in [n]$ such that
 $|\RK(x)-r| \le \delta \min(r,n-r)$. 
Given $\delta$ and $x$, the problem of finding a $\delta$-approximation for $\RK(x)$ is called the \emph{approximate rank problem}.
\label{defn:rank}
\end{definition}

Likewise, it will be useful to assume that $k \le n/2$ for selection, if this is not
the case, then we can just select the $k'=(n-k)$-th element using $\le^{\text{rev}}$ as the order,
again using the same algorithm but reversing the direction of the group tests.  We also remark that \cref{defn:rank} is stronger than
the standard notion of a $1 \pm \delta$ approximation to $\RK(x)$, since a $\delta$-approximation is tighter when $r > n/2$.

\begin{definition}
A $\delta$-approximation for the $k$-th order statistic (for $\delta \in (0,1), k \in [n]$) is an element $x \in S$ such that 
$|\RK(x)-k| \le \delta \min(k,n-k)$.  Given $\delta$ and $k$, the problem of finding a $\delta$-approximation for the $k$-th order statistic is called the \emph{approximate selection problem}.
\end{definition}

\section{Min-Finding}

We first give a Las Vegas randomized algorithm to find the minimum element in $S$.
The basic idea of the algorithm is to iteratively refine a candidate $x$ for the minimum element (initially chosen at random).  At each iteration, either $x$ is determined to be the global minimum, or an element whose rank is at most that of the previous $x$. Furthermore, the rank of the next candidate found is expected to be half of the rank of the current $x$.
The complete algorithm is given in \cref{algo:min_find}.

\begin{algorithm}[h]
\caption{Min-finding algorithm}\label{algo:min_find}
\begin{algorithmic}[1]
\Function{MinFind}{$S$}
\State $x \leftarrow \text{Randomly select an element from the set } S$.
\While{$|S| > 1$ \textbf{and} $S\setminus\{x\} \le_Q x$} 
    \State $x \leftarrow \text{Swap}(S, x)$
\EndWhile
\State \Return $x$.
\EndFunction
\Function{Swap}{$A, x$}
\If{$A=\{a\}$}  
    \State \Return $a$
\EndIf
\State $A_0, A_1 \leftarrow \text{Partition } A \text{ randomly into } A_0, A_1 \text{ s.t. } ||A_0|-|A_1|| \le 1$
\If{$A_0 \le_Q x$}
    \State \Return $Swap(A_0, x)$ 
\Else
    \State \Return $Swap(A_1, x)$ \Comment{$A_1$ must contain an element $\le x$.}
\EndIf
\EndFunction
\end{algorithmic}
\end{algorithm}

\begin{lemma}
Let $x$ be the current candidate, and let $Y = \{ y \in S : y \le x \}$.  If $|Y|>0$, then $\forall y \in Y$,
$$
\Pr[\text{Swap}(S,x)\text{ returns } y] = \frac{1}{|Y|}.
$$
\label{lem:swap1}
\end{lemma}
\begin{proof}
The lemma follows from the random partitioning of elements in the Swap function.

\end{proof}

\begin{corollary}
The expected number of iterations needed by MinFind is $\bigo(\log n)$.
\label{cor:minfind}
\end{corollary}
\begin{proof}

Let $I(n)$ be the expected number of iterations needed to find the minimum value among a set of size $n$.  We have $I(1) = 0$ and for $n > 1$:
\begin{equation}
I(n) = 1 + \frac{1}{n} \sum_{i=1}^{n} I(i),
\label{eq:recur}
\end{equation}
since $\RK(x)$ is initially uniformly distributed in $[1,n]$ and each Swap$(S, x)$ call returns an element whose rank is uniformly distributed in
$[1, \RK(x)]$ by \cref{lem:swap1}.
We can rewrite \cref{eq:recur} as:
\begin{equation}
n \cdot I(n) = n + \sum_{i=1}^{n} I(i),
\label{eq:recur2}
\end{equation}
and so for $n > 2$, we have:
\begin{equation}
(n-1) \cdot I(n-1) = n-1 + \sum_{i=1}^{n-1} I(i).
\label{eq:recur3}
\end{equation}
Subtracting \cref{eq:recur3} from \cref{eq:recur2} yields
$$
n \cdot I(n) - (n-1) \cdot I(n-1) = 1 + I(n),
$$
and so
$$
I(n) - I(n-1) = \frac{1}{n-1}.
$$
From \cref{eq:recur}, $I(2)=2$, so $I(n) = 1 + (1 + \frac{1}{2} + \frac{1}{3} + \ldots + \frac{1}{n-1})$; i.e., a shifted harmonic series, so $I(n) = \bigo(\log n)$.

\end{proof}

\begin{lemma}
Swap does $\lceil \log_2(n-1) \rceil$ group tests.
\label{lem:swap2}
\end{lemma}
\begin{proof}
The initial size of $A$ is always $n-1$ when Swap is invoked from MinFind.  
If $|A| > 1$, Swap makes one group test and then recurses on either $A_0$ or $A_1$, which are of size at most $\frac{|A|+1}{2}$.
Thus, the depth of recursion is $\lceil \log_2(n-1) \rceil$.

\end{proof}

\begin{theorem}
The min element of $S$ can be found with $\bigo(\log^2 n)$ expected group test queries.
\label{thm:minfindingtimecomplexity}
\end{theorem}
\begin{proof}
This follows directly from \Cref{cor:minfind} and \Cref{lem:swap2}.

\end{proof}
As previously mentioned, the min-finding algorithm can be converted to a max-finding algorithm by simply reversing the group test queries.

\section{Rank Determination}

We first observe that rank determination can be solved using group-testing defect counting; if we wish to determine the rank of an element $x \in S$, we can consider those elements less than or equal to $x$, i.e. $\{y \in S : y \le x \}$ as being `defective'.  The group inequality test $V \le_Q x$ is analogous to the standard group test to check if $V$ contains any defective elements. 
The algorithm from \cite{falahatgar2016estimating} can thus 
find a $1 \pm \delta$ approximation to 
$\RK(x)$, using
\(2 \log \log n + \bigo\left(\frac{1}{\delta^2}\log\frac{1}{\epsilon}\right)\)
group inequality queries, that succeeds with probability $1-\epsilon$. 
Their algorithm is based on a multistage binary search.  Below we describe a simple
binary search approach that finds a $\delta$-approximation with $\softo(\frac{1}{\delta^2} \log \frac{1}{\epsilon})$ query complexity.  

First, we introduce a useful randomized test routine that is also used later in our selection algorithm; specifically we would like to decide for any element $x$ and some $r \in [n]$, whether $\RK(x) \le r$. 
We will assume $r \le n/2$ (the case $r > n/2$ is handled by reversing the direction of the group tests and is discussed subsequently).  
We also assume that $n$ is divisible by $r$; if not, add at most $r-1$ equivalent dummy elements to $S$ that are greater than all other elements. 

We first describe the test:
Suppose $x \in S$ and $S'$ is a random sample (with replacement) of $S$ of size $N = n/r$.
If $\RK(x) \le r -\delta r$, 
\begin{align*}
\Pr[S' \le_Q x] & = 1 - \left( \frac{n - \RK(x)}{n} \right)^{N} \le 1 - \left( \frac{n - (r -\delta r)}{n} \right)^{N} \triangleq p_L.
\end{align*}
Similarly, if $\RK(x) \ge r +\delta r$, 
\begin{align*}
\Pr[S' \le_Q x] & = 1 - \left( \frac{n - \RK(x)}{n} \right)^{N} \ge 1 - \left( \frac{n - (r +\delta r)}{n} \right)^{N} \triangleq p_R.
\end{align*}
\begin{align*}
p_R - p_L & = (1 - p_L) - (1 - p_R) \\
& = \left( \frac{n - r + \delta r}{n} \right)^{N} - \left( \frac{n - r - \delta r}{n} \right)^{N} \\
& \ge \left( \frac{n - r + \delta r}{n} \right)^{N} - \left( \frac{n - r}{n} \right)^{N} \\
& = \left(1 - \frac{1}{N} + \frac{\delta}{N} \right)^{N} - \left(1 - \frac{1}{N} \right)^{N} \\
& = f(\frac{\delta}{N}) - f(0), 
\end{align*}
where $f(x) = (1 - \frac{1}{N} + x)^{N}$.  Note that
$f'(x) = N \cdot (1 - \frac{1}{N} + x)^{N-1}$ and
$f''(x) = N(N-1) \cdot (1 - \frac{1}{N} + x)^{N-2}$.  Since $N \ge 2$, $f''(x) \ge 0$ and thus $f$ is convex, and we can apply Cauchy's inequality:
\begin{align*}
p_R - p_L & \ge f(\frac{\delta}{N}) - f(0) \\
& \ge f'(0) \cdot \frac{\delta}{N} \\
& \ge  N\cdot (1 - \frac{1}{N})^{N-1} \cdot \frac{\delta}{N} \\
& > \delta \cdot (1 - \frac{1}{N})^{N} \\
& \ge \delta \cdot e^{-1} (1 - \frac{1}{N}) \ge \delta \cdot e^{-1} / 2,
\end{align*}
using the inequality $(1+\frac{x}{n})^n \ge e^x (1-\frac{x^2}{n})$ for $n \ge 1, |x| \le n$.

\begin{algorithm}[t]
\caption{Less-than rank test (for $r \le \frac{n}{2}$)}\label{algo:lessthan}
\begin{algorithmic}[1]
\Function{TestLE}{$x,r,\delta,\epsilon$}
\State $c \leftarrow 0$
\For{$i = 1$ to $N_t$} 
    \State $S' \leftarrow \text{Randomly select } n/r \text{ elements from } S \text{ with replacement}.$
    \If{$S' \le_Q x$}
        \State $c \leftarrow c + 1$.
    \EndIf
\EndFor
\State \Return $(c \ge p^* \cdot N_t)$
\EndFunction
\end{algorithmic}
\end{algorithm}
Suppose we do $N_t$ random trials of this process (i.e., pick new random $S'$ sets each trial) and let $X$ be the number of positive responses.  
Let $p^* = \frac{p_L + p_R}{2}$.  So, $p_R - p^* = p^* - p_L = \frac{1}{2} (p_R - p_L) \ge \delta/4e$.
If $\RK(x) \le r -\delta n$, by Hoeffding's inequality (since this sum has a binomial distribution),
\begin{align*}
\Pr[X \ge p^* N_t]
& = \Pr[X -p_L N_t \ge p^* N_t - p_L N_t] \\
& \le \Pr[X - p_L N_t \ge \delta/4e \cdot N_t] \le \exp(-2 (\delta/4e)^2N_t)).
\end{align*}
Similarly, if $\RK(x) \ge r +\delta n$,
\begin{align*}
\Pr[X \le p^* N_t]
& = \Pr[N_t - X \ge N_t - p^* N_t] \\
& = \Pr[(N_t - X) - (1-p_R)N_t \ge N_t - p^* N_t - (1-p_R)N_t] \\
& = \Pr[(N_t - X) - (1-p_R)N_t \ge N_t (p_R - p^*)] \\
& \le \Pr[(N_t - X) - (1-p_R)N_t \ge \delta/4e \cdot N_t] \le \exp(-2 (\delta/4e)^2N_t)),
\end{align*}
where the last inequality again follows by Hoeffding's inequality on the sum of the negative trials.
If $\epsilon > 0$ is the desired error rate, then we should set 
$\exp(-2 (\delta/4e)^2N_t)) = \epsilon$, and 
take $N_t = \frac{\ln(1/\epsilon)}{2 (\delta/4e)^2} = \frac{8e^2 \ln(1/\epsilon)}{\delta^2}$.
\Cref{algo:lessthan} implements this strategy.

Using the observation that testing $\RK(x) \le r$ for $\le$ is equivalent to
testing $\RK(x) \ge n-r$ for $\le^{\text{rev}}$, we can adapt the test to handle
the case $r > n/2$ by replacing $r$ with $n-r$ and reversing the direction of group tests.
To summarize, we have a test \textsf{TestLE}$(x,r,\delta,\epsilon)$ with the following property:
\begin{lemma}
\textsf{TestLE}$(x,r,\delta,\epsilon)$ makes $\bigo(\frac{1}{\delta^2}\ln \frac{1}{\epsilon} )$ group test queries and 
is correct with probability 
$1 - \epsilon$. 
\label{lem:testLE}
\end{lemma}

As mentioned, 
our rank determination strategy uses a binary-search approach.
We assume that $\RK(x)$ has been found to lie in an interval 
$[a -\delta \min(a,n-a), b +\delta \min(b,n-b)]$ with high probability (initially $a=1, b=n$).   Let $m = \frac{a+b}{2}$ be the midpoint of the interval. 
We call \textsf{TestLE}$(x,m,\delta,\epsilon/\lceil \log_2 n \rceil)$.  If the call returns `true', then we 
assume $\RK(x) \le m + \delta \min(m,n-m)$ and set $b=m$, otherwise we
assume $\RK(x) \ge m - \delta \min(m,n-m)$ and set $a=m$.  Each call
reduces the search range by a factor of $2$, so $\lceil \log_2 n \rceil$ calls are needed to reduce the search
range to a single value $r$, which is returned as the approximate rank.  By the union bound,
the probability that $r$ is a $\delta$-approximation to $\RK(x)$ is at least $1 - \epsilon$.
The total number of group test queries is 
\begin{align*}
N_t \cdot \lceil \log_2 n \rceil
 & = \frac{8e^2}{\delta^2} \ln(\lceil \log_2 n \rceil/\epsilon)\lceil \log_2 n \rceil \\
 & = \frac{8e^2}{\delta^2} (\ln \lceil \log_2 n \rceil + \ln 1/\epsilon) \lceil \log_2 n \rceil \\
& = \bigo(\frac{\log n}{\delta^2} (\log \log n + \log 1/\epsilon)).
\end{align*}
Let this algorithm be denoted \textsf{ApxRank}$(x,\delta,\epsilon)$ 
Thus we have,
\begin{theorem}
\textsf{ApxRank}$(x,\delta,\epsilon)$ makes $\bigo(\frac{\log n}{\delta^2} (\log \log n + \log \frac{1}{\epsilon}))$ group test
queries and returns a rank $r \in [n]$ such that $|\RK(x)-r| \le \delta \min(r,n-r)$ with 
probability $1 - \epsilon$.
\end{theorem}

\section{Selection}

In this section, we describe a Monte Carlo randomized algorithm for the approximate selection problem.  Our strategy uses two
basic steps: the first step invokes a sampling procedure to generate a candidate $x$ that is close to the desired rank with a certain probability (in this case $x$ is said to be a good candidate).
The second step uses a separate verification procedure to check that $x$ meets the approximate rank requirements.  These steps are repeated a fixed number
of times so that we have at least a $\frac{1}{2}$ chance to find a good candidate.
We will bound the query complexity and failure rates of both
steps.  
We will again assume that $n$ is divisible by $k$; if not, we will add at most $k-1$ equivalent dummy elements to $S$ that are greater than all other elements.

\subsection{Step 1: Sampling Procedure}

Let $k \le n/2$; as before, the case $k \ge n/2$ can be handled by reversing the direction
of the group tests in the algorithm and then selecting the $n-k$-th element.
We start by analyzing the chance that the minimum of a sample
of size $N = n/k$ falls in the range $(k-\delta k, k+\delta k]$.

\begin{lemma}
For $0 < k \le (1-e^{-2\delta}) n/2$, 
let $m$ be the minimum element of a randomly sampled (with replacement) subset from $S$ of size $N$.  Then,
\begin{equation*}
\Pr[k - \delta k < \RK(m) \le k + \delta k] \ge \frac{\delta^2}{4}.
\end{equation*}
\label{lem:smalldelta}
\end{lemma}
\begin{proof}
Note that 
\begin{align*}
\Pr[\RK(m) > k + \delta k] 
 & = \left( \frac{n - (k+\delta k)}{n} \right)^{N} \\
 & = \left(1 - \frac{(1 + \delta )}{N}\right)^{N}\\
 & \le e^{-(1+\delta)}
\end{align*}
using the fact that $e^x \ge (1  + \frac{x}{n})^n$ for $n \ge 1$.
Next, observe
\begin{align}
\Pr[\RK(m) \le k - \delta k]
 & = 1 - \left( \frac{n - (k-\delta k)}{n} \right)^N \nonumber \\
 & = 1 - \left(1 - \frac{(1 -\delta)}{N}\right)^{N} \nonumber\\
 & \le 1 - e^{-(1-\delta)}\left(1 - \frac{(1-\delta)^2}{N}\right) \label{eq:elower}\\
 & \le 1 - e^{-(1-\delta)}\left(1 - \frac{1}{N}\right) \nonumber\\
 & \le 1 - e^{-(1-\delta)}\left(1 - \frac{1-e^{-2\delta}}{2}\right)  \label{eq:N} \\
 & = 1 - e^{\delta-1}\left(\frac{1+e^{-2\delta}}{2}\right)  \nonumber
\end{align}
where \cref{eq:elower} 
uses $(1+x/n)^n \ge e^x(1-x^2/n)$ when $n \ge 1, |x| \le n$ and
\cref{eq:N} uses the fact that $k \le (1-e^{-2\delta}) n/2$ implies $\frac{1}{N} \le \frac{1-e^{-2\delta}}{2}$.  
Combining both bounds, we find
\begin{align}
\Pr[k- \delta k < \RK(m) \le k + \delta k] 
 & = 1 - \Pr[\RK(m) \le k - \delta k] - \Pr[\RK(m) > k + \delta k]\nonumber\\
 & \ge e^{\delta-1}\left(\frac{1+e^{-2\delta}}{2}\right) - e^{-(1+\delta)} \nonumber\\
 & = e^{\delta-1} \cdot \left( \left( \frac{1+e^{-2\delta}}{2}\right) - e^{-2\delta} \right) \nonumber\\
 & = \frac{e^{\delta-1}}{2} (1-e^{-2\delta}) \nonumber\\
 & \ge \frac{\delta}{2} (1-e^{-2\delta}) \label{eq:elower2} \\
 & > \frac{\delta^2}{4}, \label{eq:delta}
\end{align} 
where \cref{eq:elower2} follows from $e^x \ge 1  + x$.
Since $e^{-x} \le 1-\frac{x}{2}$ for $x \in [0, 1.59]$, 
so $1 - e^{-2\delta} \ge 1 - (1-\frac{2\delta}{2})=\delta$, when $\delta \le \frac{1}{2}$.  
For $\frac{1}{2} < \delta < 1$, $1 - e^{-2\delta} \ge 1-e^{-1} > \frac{\delta}{2}$.  
So, $1 - e^{-2\delta} \ge \frac{\delta}{2}$, used in \cref{eq:delta} and \cref{eq:deltaagain} below.

\end{proof}

\begin{lemma}
For $(1-e^{-2\delta}) n/2 < k \le n/2$, 
let $x$ be a single random sample from $S$.  Then,
\begin{equation*}
\Pr[k- \delta k < \RK(x) \le k + \delta k] > \frac{\delta^2}{4}.
\end{equation*}
\label{lem:largedelta}
\end{lemma}
\begin{proof}
\begin{align}
\Pr[k - \delta k < \RK(x) \le k + \delta k] 
& \ge \frac{ \delta k }{n} \nonumber\\
& > \delta \frac{(1-e^{-2\delta})}{2} \nonumber\\
& \ge \frac{\delta^2}{4}. \label{eq:deltaagain}
\end{align}

\end{proof}

\begin{algorithm}[t]
\caption{Random sample - Step 1}\label{algo:min_point_selection}
\begin{algorithmic}[1]
\Function{GetCandidate}{$k, \delta$}
\If{$k \le (1-e^{-2 \delta}) n/2$}
\State $S' \leftarrow \text{Randomly select } \frac{n}{k} \text{ elements from } S \text{ with replacement}.$
\State $x \leftarrow \text{MinFind}(S')$
\Else
\State $x \leftarrow \text{Randomly select an element from $S$.}$
\EndIf
\State \Return $x$
\EndFunction
\end{algorithmic}
\end{algorithm}

\Cref{algo:min_point_selection} summarizes the sampling procedure.  
Let $I$ be the number of iterations needed to first draw a $\delta$-approximation to the $k$-th order statistic.
From \Cref{lem:smalldelta} and \Cref{lem:largedelta}, we have $\EX(I) \le 4/\delta^2$.
Furthermore, $\Pr[I \ge 2 \EX(I)] \le \frac{1}{2}$, by Markov's inequality.  Thus, if we draw 
$8/\delta^2$ samples, we will find a good sample with probability at least $\frac{1}{2}$.
In fact, it will be useful to have a $1/2$ chance to get a $\delta/2$-approximation, so instead we will draw $32/\delta^2$ samples.

\subsection{Step 2: Checking a Sample}

Next we need to check whether the sample $x$ from Step 1 is a $\delta/2$-approximation to the $k$-th order statistic.
We can use the rank comparison test, \textsf{TestLE} (\cref{algo:lessthan}), from the previous section to check $x$:
Specifically \textsf{TestLE}$(x,k+\frac{3}{4}\delta k,\delta',\epsilon')$ can check that
$\RK(x) \le k + \frac{1}{2} \delta k$ if we take 
$\delta' = \frac{\delta k/4}{k+\frac{3}{4}\delta k} = \frac{\delta/4 }{1+\frac{3}{4}\delta} > \delta/8$; this
will be reliable with probability $1 - \epsilon'$.
Similarly,
\textsf{TestLE}$(x,k-\frac{3}{4}\delta k,\delta'',\epsilon')$ can check that
$\RK(x) \ge k - \frac{1}{2} \delta k$, if we take 
$\delta'' = \frac{\delta k/4}{k-\frac{3}{4}\delta k} = \frac{\delta/4}{1-\frac{3}{4}\delta} > \delta/4$; this
will again be reliable with probability $1 - \epsilon'$.
If both of these inequality tests pass, then we accept $x$.
We have:
\begin{lemma}
Let $x$ be a randomly sampled element from Step 1.
If $|\RK(x) - k| \le \delta/2 \min(k, n-k)$, then $x$ is accepted with probability $1-2\epsilon'$, and if $|\RK(x) - k| > \delta \min(k, n-k)$, then $x$ is rejected with probability $1-\epsilon'$.
\end{lemma}
\begin{proof}
If $|\RK(x) - k| \le \delta/2 \min(k, n-k)$, then by \cref{lem:testLE}, each inequality test is 
passed with probability $1-\epsilon'$.  By the union bound, both are passed with probability
$1-2\epsilon'$.  Suppose $|\RK(x) - k| > \delta \min(k, n-k)$.  If $\RK(x) > k$ then 
\textsf{TestLE}$(x,k+\frac{3}{4}\delta k,\delta',\epsilon')$ returns false with probability
$1-\epsilon'$.  Similarly, if $\RK(x) < k$, \textsf{TestLE}$(x,k-\frac{3}{4}\delta k,\delta'',\epsilon')$ will return true with probability $1-\epsilon'$.  Thus, $x$ is
correctly rejected with probably $1-\epsilon'$.

\end{proof}
If $x$ is accepted, we report it as the selected element; if it is rejected, then another 
sample is drawn.  
The number of Step 2 checks performed is at most $32/\delta^2$; any sample that should be rejected is rejected with probability $1-\epsilon'$.  If we draw a sufficiently good sample (a $\delta/2$-approximation) it is accepted with probability $1-2\epsilon'$.
Thus, by the union bound, we should set 
$\epsilon' = \frac{1}{32/\delta^2 + 1} \cdot \epsilon$ 
to obtain an overall $\epsilon$
probability of failure (all bad elements are rejected and a good one is accepted, if drawn). 
We denote the full
approximate selection algorithm as \textsf{ApxSelect}$(k,\delta,\epsilon)$ (see \cref{alg:apxselect}).
\begin{algorithm}[t]
\caption{Full selection algorithm}\label{alg:apxselect}
\begin{algorithmic}[1]
\Function{ApxSelect}{$k, \delta, \epsilon$} 
\For{$i = 1$ to $32/\delta^2$}
\State $x \leftarrow \text{GetCandidate}(k, \delta/2)$

\If{\textsf{TestLE}$(x,k+\frac{3}{4}\delta k,\delta',\epsilon')$  \textbf{and not} \textsf{TestLE}$(x,k-\frac{3}{4}\delta k,\delta'',\epsilon')$ }
\State \Return $x$
\EndIf
\EndFor
\State \Return not found
\EndFunction
\end{algorithmic}
\end{algorithm}
We have,
\begin{theorem}
\textsf{ApxSelect}$(k,\delta,\epsilon)$ makes an expected 
$\bigo(\frac{1}{\delta^2}\log^2 N + \frac{1}{\delta^4}\log \frac{1}{\epsilon \delta^2})$ group test
queries and with probability at least $\frac{1}{2}$ returns an element $x \in S$ 
such that $|\RK(x)-k| \le \delta \min(k,n-k)$ with probability $1 - \epsilon$.
\end{theorem}
\begin{proof}
As discussed above, if the algorithm outputs an element $x$, then it is a $\delta$-approximation with probability $1-\epsilon$, and there is at least a $1/2$ chance that this occurs.
The number of iterations of the outer loop is at most $32/\delta^2$. 
Each call to \textsf{GetCandidate} makes an expected $\bigo(\log^2 N)$ queries.
Each call to \textsf{TestLE} makes at most
$$
\frac{8e^2 \ln(1/\epsilon')}{(\delta/8)^2} = \bigo(\log(\frac{1}{\epsilon \delta^2})/\delta^2)
$$
queries.  The total expected number of queries is thus
$$
\bigo(\frac{1}{\delta^2} (\log^2 N + \log(\frac{1}{\epsilon \delta^2})/\delta^2)).
$$

\end{proof}

\section{Discussion}

In this work, we considered several classic problems on order-finding where a group inequality testing operation is available; we gave randomized algorithms for min-finding, rank determination, and selection. A natural question is whether group inequality testing might also be useful for similar problems on partially-ordered sets.  For posets, there can be multiple minimum elements (up to the \emph{width}\footnote{The size of a maximum antichain in the poset; i.e., the maximum number of mutually incomparable elements.} of the poset.).  \Cref{algo:min_find} will find one, but there does not seem to be an efficient way to determine all minimum elements (or estimate the width of the poset).  In addition, the rank of an element $x$ of a poset is normally defined as the length of the longest left-extending chain from $x$, not the number of elements less than or equal to the element.  Thus, the approach used in \cref{algo:lessthan} cannot be directly applied as the number of elements less than or equal to an element of rank $r$ can vary between $r$ and $wr -w + 1$, where $w$ is the width of the poset.
We note that the trivial lower bound of $\Omega(\log n)$ applies to each of the problems
considered, so it is possible that non-trivial lowers bounds might be found, e.g. for min-finding with group testing.
Furthermore, a more powerful group test of the form $U \le_Q V$ may be considered, although it does not seem to benefit the algorithms presented.  Another variation is a ``noisy'' version of the group test in which there is some probability to get the wrong answer from the oracle. In this case, the oracle can be resampled to decrease the error rate; the algorithms here can all be adapted to this setting, but we leave this as future work.

\subsubsection*{Acknowledgments} This work has received funding from US National Science Foundation awards 2309902 and 2243010.

\bibliography{main}

\end{document}